\documentclass[final,twocolumn,3p,times]{elsarticle}
\usepackage{amssymb}
\usepackage{amsmath}
\usepackage{physics}
\usepackage{xcolor}

\journal{Physics Letters A}

\begin{document}

\begin{frontmatter}



\title{Quantum thermalization and average entropy of a subsystem} %% Article title


\author{Smitarani Mishra} %% Author name
\author{Shaon Sahoo \corref{cor1}} %% Author name
\ead{shaon@iittp.ac.in}
\affiliation{organization={Department of Physics, Indian Institute of Technology Tirupati},%Department and Organization
             addressline={Yerpedu}, 
%             city={Tirupati},
            postcode={517619}, 
%            state={Andhra Pradesh},
            country={India}}

\cortext[cor1]{Corresponding author}

%% Abstract
\begin{abstract}
Page's seminal result on the average von Neumann (VN) entropy does not immediately apply to realistic many-body systems which are restricted to physically relevant smaller subspaces. We investigate here the VN entropy averaged over the pure states in the subspace $\mathcal{H}_E$ corresponding to a narrow energy shell centered at energy $E$. We find that the average entropy is  $\overline{S}_{1} \simeq \ln d_1$, where $d_1$ represents first subsystem's effective number of states relevant to the energy scale $E$. If $d_E = \dim{(\mathcal{H}_E)}$ and $D$ ($D_1$) is the Hilbert space dimension of the full system (first subsystem), we estimate that $d_1 \simeq D_1^\gamma$, where $\gamma = \ln (d_E) / \ln (D)$ for nonintegrable (chaotic) systems and $\gamma < \ln (d_E) / \ln (D)$ for integrable systems. This result can be reinterpreted as a volume-law of entropy, where the volume-law coefficient depends on the density-of-states for nonintegrable systems, and remains below the maximal possible value for integrable systems. 
We numerically analyze a spin model to substantiate our main results.
\end{abstract}

%% Keywords

\end{frontmatter}

\section{Introduction} \label{sec1}
Understanding how thermodynamic behavior emerges from the microscopic rules of quantum mechanics remains a central question in statistical physics \cite{Gogolin-Eisert16,DAlessio-Rigol16,Reimann08}. One particularly fruitful avenue of investigation concerns the entanglement structure of many-body quantum systems, especially in the context of thermalization \cite{Goldstein06,Popescu06}. When a large quantum system is in a pure state, the entanglement entropy of its small subsystem often mimics the thermodynamic entropy \cite{Sharma13, Goldstein19, Landau_SP,Mori18,Deville13, Peres_QTCM,Neumann_MFQM}, suggesting a deep connection between quantum entanglement and statistical mechanics.

The von Neumann (VN) entropy quantifies the bipartite quantum entanglement of a pure state \cite{Horodecki09}. This VN entropy of typical pure states provides a key insight into the statistical thermodynamical properties of the quantum systems.  For example, the VN entropy of typical eigenstates of quantum many-body Hamiltonians may help us determine whether a system is integrable or chaotic (nonintegrable) \cite{Bianchi22,Leblond19}; this in turn determines how the system thermalizes \cite{DAlessio-Rigol16,Nandy16,Cassidy11}. In a seminal work, Page first provided an exact formula for the VN entropy of a Haar-random pure state \cite{Page93}. Later, this formula was rigorously proved by others \cite{Sen96,Foong94,Ruiz95}. In another important work, Vidmar and coworkers reported the exact bounds for the VN entropy of typical eigenstates (for translationally invariant quadratic fermionic Hamiltonians \cite{Vidmar17}). Many important works subsequently reported the VN entropy of typical pure states from different ensembles or of typical eigenstates of physical Hamiltonians \cite{Bianchi22,Bianchi21,Bianchi19,Fukuda19,Garrison18}.

In this work, we investigate the average VN entropy of a subsystem, $\overline{S}_{1}$, when the full quantum system is constrained to a physically relevant subspace corresponding to a narrow energy shell- in other words, when the full system resides in a random pure state with a fixed energy scale. Our study is motivated by the following objectives:
\begin{enumerate}
\item To provide a complementary understanding or alternative perspective of quantum thermalization, which is traditionally understood through frameworks such as canonical typicality \cite{Goldstein06,Popescu06} and the Eigenstate Thermalization Hypothesis or ETH \cite{DAlessio-Rigol16,Deutsch91,Srednicki94}. 
\item To explore how the average subsystem entropy $\overline{S}_{1}$ scales with the density of states (DOS) of the full system. This result offers a compelling connection between the VN entropy and the thermodynamic (Boltzmann) entropy of an equilibrium quantum system. 
To the best of our knowledge, Ref. \cite{Sahoo12} provided the first numerical evidence that the average VN entropy over a subspace scales logarithmically with the DOS at the corresponding energy, i.e., $\overline{S}_{1} \propto \ln{(\mathrm{DOS})}$. 
\item To examine the scaling of average VN entropy $\overline{S}_{1}$ with the subsystem size $l_1$. It is well established that highly excited states typically obey a volume law for entanglement entropy \cite{Molter14,Alba09,Moudgalya18,Bianchi22}. The value of the proportionality coefficient in the volume law can serve as an indicator of the system's dynamical character - whether the system is nonintegrable (chaotic) or integrable \cite{Leblond19}. 
\end{enumerate}

We present a simple analytical model to compute the average von Neumann entropy $\overline{S}_{1}$ associated with a narrow energy shell, focusing on the leading-order contribution and neglecting subleading corrections. As we shall see, this leading term already captures the essential features relevant to the questions addressed in this work. Our main result is summarized as follows.

When the full system is restricted to a specific narrow energy window, the average VN entropy $\overline{S}_{1}$ of the (first) subsystem follows a volume-law scaling. However, the coefficient of this law is not constant; rather, it depends on the energy scale through the density of states (DOS), as expressed in Eq. \eqref{s1_lndos2}. We argue that this coefficient in volume-law is universal and attains its maximal value in nonintegrable (chaotic) systems, while in integrable systems it is nonuniversal and takes smaller values.

The significance of this result lies in its potential as a diagnostic tool for distinguishing between nonintegrable and integrable systems with reduced computational effort. Unlike Page’s result, which is system-independent, earlier studies by Vidmar, Rigol, and collaborators indicate that the VN entropy averaged over all eigenstates can differentiate between nonintegrable and integrable systems \cite{Leblond19,Lydzba20}. However, applying that approach to a given many-body quantum system requires diagonalization of the full Hamiltonian - a computationally intensive task. By contrast, our method relies only on eigenstates within a narrow energy shell, which can be obtained efficiently using techniques such as the shift-and-square method \cite{Sahoo19}.

To substantiate our main result, we also perform numerical calculations on a one-dimensional spin-1/2 chain with next-nearest neighbor interactions. 

\section{Average entropy at an energy scale} \label{sec2}

If a quantum system of Hilbert space dimension $D_1D_2$ is in a Haar-random pure state, then the average VN entropy of the subsystem of Hilbert space dimension $D_1$ is
\begin{equation}
\mathbb{E}({S}_{1}) = \left(\sum_{k=D_2+1}^{D_1D_2}\frac{1}{k}\right)-\frac{D_1-1}{2D_2}.
    \label{page_main}
\end{equation}
This general result was first conjectured by Page \cite{Page93} and later proven by others \cite{Sen96,Foong94,Ruiz95}. In the thermodynamic limit (with $1\ll D_1\ll D_2$), it becomes
\begin{equation}
\mathbb{E}({S}_{1}) \simeq \ln(D_1).
    \label{page_th}
\end{equation}
 Here we outline a method for calculating the average VN entropy (main term) corresponding to a subspace associated with an energy scale. We argue that, in this case, a result analogous to Eq. \ref{page_th} can be obtained. 

 \subsection{Average entropy in terms of a parameter $d_1$}
 Consider the subspace spanned by the eigenkets of the full system corresponding to a narrow energy shell: $\mathcal{H}_E=\mathrm{Span}\left\{\ket{n} : E_n\in (E+\frac{\Delta E}{2},E-\frac{\Delta E}{2} ]\right\}$, where $E_n$ is the eigenvalue associated with the eigenket $\ket{n}$ of the full system. The shell width $\Delta E$, is arbitrary, but is macroscopically small and microscopically large. A general state in the subspace can be written as $\ket{\psi}=\sum'_{n}c_n\ket{n}$ with $\sum'_n |c_n|^2=1$. Here $\sum'_n$ denotes a summation over $d_E$ energy eigenkets in $\mathcal{H}_E$, where $d_E=\dim{(\mathcal{H}_E)}$. The reduced density matrix for the subsystem 1 is 
\begin{equation}\label{rho1}
\rho_1=\mathrm{Tr}_2(\ket{\psi}\bra{\psi})={\sum_{n,m}}'c_nc^*_m\mathrm{Tr}_2(\ket{n}\bra{m}),\end{equation}
where the trace is taken with respect to a basis of the subsystem 2. When the full system is in pure state $\ket{\psi}$, the VN entropy of the subsystem 1 is given by 
\begin{equation}\label{vne1}
S_1(\rho_1)=-\mathrm{Tr}_1(\rho_1 \ln{\rho_1}),
\end{equation}
where the trace is taken with respect to a basis of the subsystem 1. To obtain $\overline{S}_{1}$, i.e., the average VN entropy corresponding to the subspace $\mathcal{H}_E$, we need to consider all possible states $\ket{\psi}$ and associated reduced density matrices $\rho_1$ (see Eq. \ref{rho1}). This, in principle, can be done by integrating $S_1(\rho_1)$ over the surface of the $2d_E$-dimensional unit sphere in the coefficient space (there are $d_E$ coefficients $c_n$, each having real and imaginary parts). If $dK$ is the surface element on the unit sphere, we have: 
\begin{equation}\label{avne1_0}
    \overline{S}_{1}=\frac{\int S_1(\rho_1)dK}{\int dK}.
\end{equation}
Performing this calculation can be challenging. One can somewhat simplify this calculation by noting that the leading term in the average VN entropy when averaged over random pure states and when averaged over energy eigenstates are the same \cite{Page93,Vidmar17}. This suggests that, for calculating the leading term, we can replace Eq. \ref{avne1_0} by the following average: 
\begin{equation}\label{avne1_1}
\overline{S}_{1}\simeq\frac{1}{d_E}{\sum_n}' S_1(\rho_1^{(n)}),
\end{equation}
where $\rho_1^{(n)}=\mathrm{Tr}_2(\ket{n}\bra{n})$. We recall that $d_E=\dim{(\mathcal{H}_E)}$. Evaluating even Eq. \ref{avne1_1} is not straightforward. Since $S_1\le \ln{(D_1)}$, we can in general write Eq. \ref{avne1_1} as
 \begin{equation}
 \overline{S}_{1} \simeq \ln d_1,
    \label{avne1_2}
\end{equation}
where $d_1$ is some parameter and, clearly, $d_1\le D_1$. Primary objective of our study is to find the value of $d_1$.

\subsection{Physical interpretation of the parameter $d_1$}
To obtain a physical interpretation of $d_1$, we note the following. When we randomly choose pure states (of the full system) without restriction on the energy scale $E$, then all the energy scales associated with subsystem 1 are important. Accordingly, all states ($D_1$ in number) of subsystem 1 contribute to its average VN entropy. This is an intuitive way to understand the result in Eq. \ref{page_th}.
When attention is restricted to the pure states in the subspace $\mathcal{H}_E$ associated with a narrow energy shell centered at energy $E$, all the states of subsystem 1 are no more equally significant. The parameter $d_1$ in Eq. \ref{avne1_2} can be interpreted as the effective number of significant (orthonormal) states of subsystem 1 when the full system is restricted to the subspace $\mathcal{H}_E$. We can also view $d_1$ as the dimension of the first subsystem's effective Hilbert space associated with the subspace $\mathcal{H}_E$.

To better understand $d_1$, we consider a thermodynamically large system with $1\ll l_1,l_2$, where $l_1$ ($l_2$) is the size of subsystem 1 (subsystem 2). Here, the size of a subsystem can simply be taken as its number of sites. In the thermodynamic limit, we ignore the interaction energy between the subsystems to calculate the leading term in $\overline{S}_{1}$. This can be justified when the system is large and the Hamiltonian is local (i.e., interactions are local). It is expected that, for the large system considered here, the states of subsystem 1 corresponding to the energy scale $\overline{E}_1=\frac{l_1}{l_1+l_2} E$ would be the most important. Thus the parameter $d_1$ represents the effective number of (orthonormal) states of subsystem 1 corresponding to its energy scale $\overline{E}_1$.

\subsection{Understanding subsystem energy scale}
It is possible to intuitively understand why the states of subsystem 1 corresponding to the energy scale $\overline{E}_1=\frac{l_1}{l_1+l_2} E$ would be the most important when the full system is at the energy scale $E$ (fixed). In the following, we first explain this (semi-) classically, then, elucidate it quantum mechanically using the perturbation theory. 

If the full system is restricted in the energy shell centered at energy $E$ (fixed), then $E_1+E_2=E$. Here $E_1$ ($E_2$) is the energy of the subsystem 1 (subsystem 2). We note that, although the energy of the full system is almost fixed, the energy of a subsystem is not. For example, for the subsystem 1, $0<E_1<E$. To know the most probable value of $E_1$, one needs to look at the probability density function or PDF corresponding to subsystem 1. If $p_1(E_1)$ denotes the PDF, it can be shown that \cite{Pathria_SM}: \begin{equation}\label{prob_fn}
p_1(E_1)\propto e^{-\frac{(E_1-U_1)^2}{2k_BT^2C_1}},
\end{equation} 
where $k_B$ is the Boltzmann constant, $T$ is the equilibrium temperature of the system and $C_1$ is the specific heat at constant volume (for the subsystem 1). In the expression of $p_1(E_1)$, $U_1$ is the thermodynamic internal energy of subsystem 1. Since the energy of the full system is approximately $E$, its thermodynamic internal energy would be $U\simeq E$. Since $U$ is an extensive quantity, the internal energy corresponding to subsystem 1 would be $U_1=\frac{l_1}{l_1+l_2}U\simeq \frac{l_1}{l_1+l_2}E = \overline{E}_1$ (here, we ignore the interaction energy between the subsystems). It is clear from Eq. \ref{prob_fn} that, the PDF is a Gaussian distribution centered at $E_1=U_1 (\simeq \overline{E}_1)$. If the size of subsystem 1 is large, then the peak will be very sharp (similar to the delta function). This result explains why it is expected that the states of subsystem 1 corresponding to the energy scale $\overline{E}_1$ would be the most significant in the thermodynamic limit.

To understand the energy scale of subsystem 1 using the perturbation theory, let us consider that the total Hamiltonian of the full system is $H_T = H_1+H_2+H_I$, where $H_1$ ($H_2$) is the Hamiltonian of the subsystem 1 (subsystem 2), and $H_I$ represents the interactions between the subsystems. Let $\ket{1,n}$ ($\ket{2,m}$) be an eigenket of $H_1$ ($H_2$) with an eigenvalue $E_{1,n}$ ($E_{2,m}$). Without the interaction term, the tensor product of two eigenkets, one for each subsystem, will be an eigenket of $H_T$. For further discussion, we adopt the following notation: $E_{n,m}=E_{1,n}+E_{2,m}$ and $\ket{n,m}=\ket{1,n}\otimes\ket{2,m}$. Consider now the following set of tensor product states: $\mathcal{T}_E=\left\{\ket{n,m}: E_{n,m}\in (E+\frac{\Delta E}{2},E-\frac{\Delta E}{2} ]\right\}$. All the kets in $\mathcal{T}_E$ are zeroth-order eigenkets of $H_T$. One can use it as a basis to represent any state of the full system at the energy scale $E$. Using perturbation theory, one gets the following eigen energies and eigenkets with the first order corrections \cite{Sakurai_MQM}: 
\begin{equation}\label{prtrb}
    \begin{split}
        \widetilde{E}_{n,m}&=E_{n,m}+\bra{n,m}H_I\ket{n,m}\\
        \widetilde{\ket{n,m}}&=\ket{n,m}+{\sum_{k,l}}'\frac{\bra{k,l}H_I\ket{n,m}}{E_{n,m}-E_{k,l}}\ket{k,l},
    \end{split}
\end{equation}
 where the summation $\sum'_{k,l}$ is over all the states in $\mathcal{T}_E$ except $\ket{n,m}$. Of-course, this result makes sense only when $|\bra{k,l}H_I\ket{n,m}|\ll |E_{n,m}-E_{k,l}|$. We note that the first order eigenket can be written as a linear combination of the basis kets in the follwoing way: $\widetilde{\ket{n,m}}=\ket{n,m}+\sum'_{k,l} c_{k,l}\ket{k,l}$, where $c_{k,l}=\frac{\bra{k,l}H_I\ket{n,m}}{E_{n,m}-E_{k,l}}$.

To assess the relevance of an eigenket $\ket{1,k}$ of subsystem 1 in constructing an eigenket of the full system approximately at energy $E$, we set \(E_{n,m} = E = \overline{E}_1 + \overline{E}_2 .\) With this, the denominator in Eq. \ref{prtrb} can be expressed as \( (\overline{E}_1 - E_{1,k}) + (\overline{E}_2 - E_{2,l}).\) Now, if $\lvert \overline{E}_2 - E_{2,l} \rvert$ is kept fixed, an increase in $\lvert \overline{E}_1 - E_{1,k} \rvert$ generally reduces the magnitude of the coefficient $c_{k,l}$. Moreover, for systems with only short-range interactions, the interaction Hamiltonian $H_I$ is diagonally dominated in the product basis. Consequently, unless $\lvert n-k \rvert$ is small, or equivalently, unless $\lvert \overline{E}_1 - E_{1,k} \rvert$ is small, the matrix element $\bra{k,l}H_I\ket{n,m}$ (and thus $\lvert c_{k,l}\rvert$) is expected to be negligibly small.

Since, a small value of $\lvert c_{k,l} \rvert$ implies that the corresponding eigenket $\ket{1,k}$ of subsystem 1 contributes insignificantly to the eigenket of the full system at energy $E$, one can conclude here that the subsystem 1 states with energies near $\overline{E}_1$ are the most important when the full system is at the energy scale $E$.
   
\subsection{Estimation of $d_1$ when $l_1\sim l_2$}
Here we estimate the parameter $d_1$ assuming that the sizes of the subsystems are of the same order, that is, $l_1\sim l_2$. For a thermodynamically large system, the average VN entropy and the VN entropy of a typical pure state are the same in the leading order \cite{Bianchi19}. Thus we estimate the average entropy by calculating the entropy of a typical pure state. In the following, we first reinterpret Eq. \ref{avne1_2} as the VN entropy of a typical state, i.e. $\overline{S}_1\simeq S_1 \simeq \ln {d_1}$, and then we find the parameter $d_1$ for the typical state.

As mentioned in the previous subsection, any state of the full system at the energy scale $E$ can be represented by a linear combination of kets in $\mathcal{T}_E$. Accordingly, we write a typical state at energy scale $E$ in the following way:
\begin{equation}\label{ket_typ}
    \ket{\psi}={\sum_{n,m}}'c_{n,m}\ket{n,m},
\end{equation}
where the summation $\sum'_{n,m}$ is over tensor product states in $\mathcal{T}_E$. 
Here, $c_{n,m}$ are random coefficients sampled from, say, a Gaussian distribution with zero mean, subject to the normalization constraint $\sum'_{n,m} |c_{n,m}|^2 = 1$.
The elements of the reduced density matrix of subsystem 1 are given by:
\begin{equation} \label{rho1_elm}
    (\rho_1)_{k,l}=\sum_{m=1}^{d_2}c_{k,m}c^*_{l,m},
\end{equation}
where $d_2$ is the number of (orthonormal) kets of subsystem 2 at the energy scale $\overline{E}_2=\frac{l_2}{l_1+l_2}E$. If we take the random numbers $c_{n,m}$ to be of the same order, i.e., $|c_{n,m}|\sim\frac{1}{\sqrt{d_1d_2}}$, then all the diagonal elements of $\rho_1$ would be of the order $(\rho_1)_{i,i}\sim \frac{1}{d_1}$. The off-diagonal elements of $\rho_1$ will be vanishingly small in the thermodynamic limit. This shows that the VN entropy of a typical state at the given energy scale $E$ is $S_1=-\text{Tr}{(\rho_1\ln{\rho_1})}\simeq \ln{d_1}$, similar to Eq. \ref{avne1_2}. 

Now, to find $d_1$, we note that if $d_1$ ($d_2$) is the number of (orthonormal) states of the subsystem 1 (subsystem 2) at the energy scale $\overline{E}_1$ ($\overline{E}_2$), and if $d_E$ is the number of eigenstates of the full system at the energy scale $E$, then 
\begin{equation}
  d_1 d_2 \simeq d_E.
    \label{eq_de}
\end{equation}
This relation is expected to hold in the case of large nonintegrable (chaotic) systems. By contrast, for integrable systems the presence of additional conserved quantities imposes constraints, leading to \(d_1 d_2 < d_E\), since the full system cannot dynamically explore the entire subspace $\mathcal{H}_E$ (we elaborate more on this point in the next subsection). When the subsystem sizes are comparable, $l_1 \sim l_2$, it follows that $d_1 \sim d_2$. Using Eq.\ref{eq_de}, we then obtain \(
d_1^2 \simeq d_E\). Substituting this into Eq. \ref{avne1_2}, we arrive at the result 
$\overline{S}_1 \simeq \tfrac{1}{2}\ln d_E$  
for a nonintegrable system.
Since $d_E\propto$DOS, the leading term in the average VN entropy becomes:
\begin{equation}
  \overline{S}_1\simeq \frac 12 \ln{(\textrm{DOS})}.
    \label{s1_lndos}
\end{equation}
For integrable system, the the average VN entropy would be less that the maximum possible value: $\overline{S}_1 < \frac 12 \ln{(\textrm{DOS})}$.

\subsection{Estimation of $d_1$ when $l_1\ll l_2$} \label{sec2.5}
Discussion in the previous subsection reveals how the average VN entropy scales with the DOS, but does not tell us how it scales with the subsystem size $l_1$. For this we consider the limit where the subsystem sizes are quite different. In the following, we consider the thermodynamic limit with $1 \ll l_1\ll l_2$.

Since $d_1\ll d_2$, following our assumption that $l_1\ll l_2$, the Schmidt decomposition of a typical state (Eq. \ref{ket_typ}) will have $d_1$ non-zero terms. Consequently, we expect $S_1\simeq \ln{d_1}$ in the current case also. We present in the following a scaling analysis to find $d_1$. 

\begin{figure*}
    \centering
    \includegraphics[width=12.2cm, height=6.1cm]{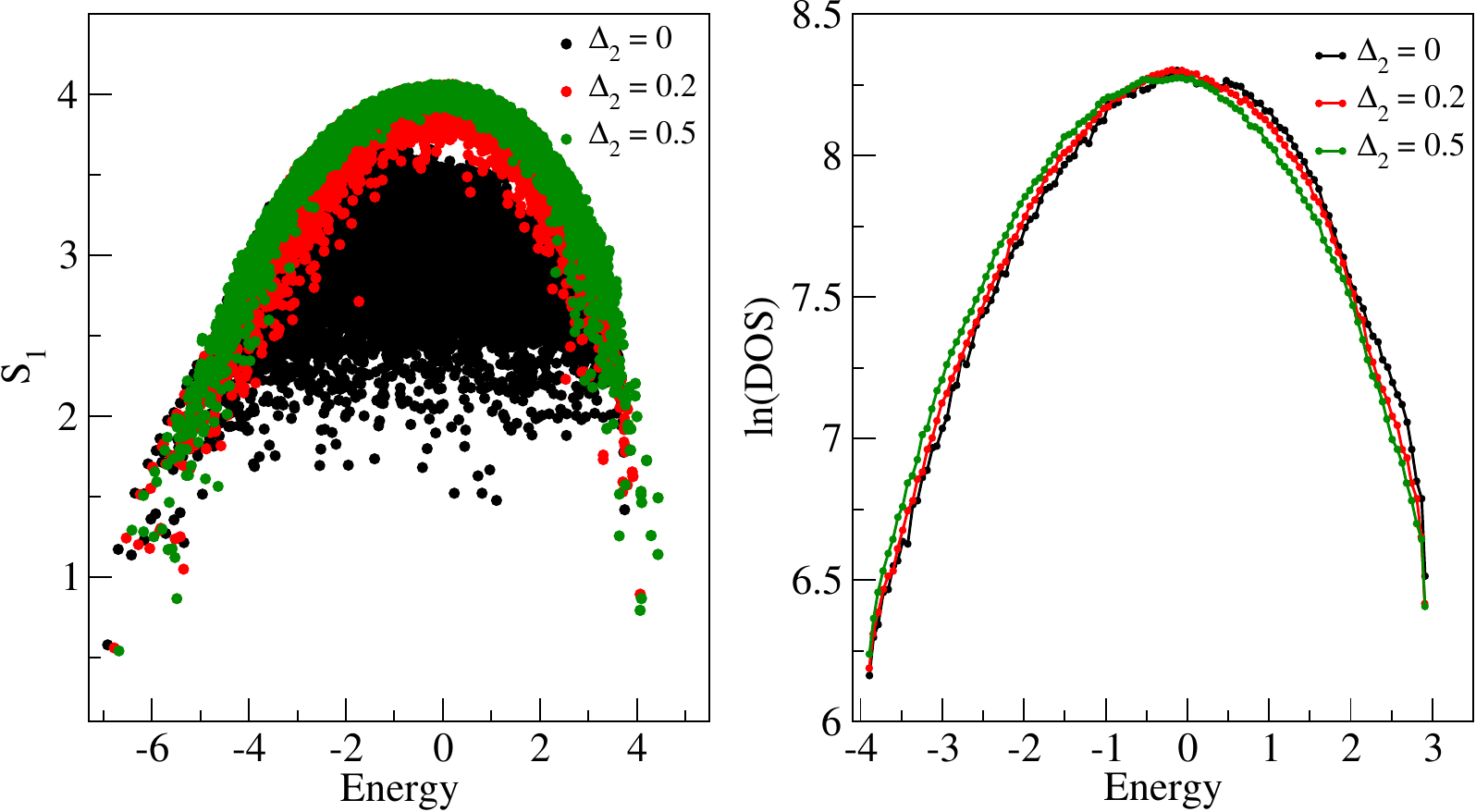}
    \caption{The subsystem VN entropy ($S_1$) for the individual eigenkets and $\ln$(DOS) are plotted against the energy of the full system (described by the Hamiltonian of Eq. \ref{spn_ham}). Calculations are performed with full system size $N=16$ and subsystem size $l_1=6$.}
    \label{ee_lndos_eng}
\end{figure*}

We now consider a parameter $\eta$ that may be called the effective Hilbert space dimension associated with a single site. The parameter is taken in such a way that $\eta^{l_1}=d_1$ and $\eta^{l_2}=d_2$. This is similar to $d^{l_1}=D_1$ and $d^{l_2}=D_2$ where $d$ is the Hilber space dimension of a single spin (for spin-1/2 system, $d=2$). This shows that 
\begin{equation}
 \frac{\ln d_1}{\ln d_2}=\frac{\ln D_1}{\ln D_2}.
    \label{eq_d1byd2}
\end{equation}
We here assume that the Eq. \ref{eq_de} is valid for nonintegrable systems even when $l_1\ll l_2$. 
Now starting from Eq. \ref{eq_de}, and using Eq. \ref{eq_d1byd2}, we get, 
\begin{equation}\label{ln_de}
\ln d_E \simeq \ln d_1(1+ \frac{\ln d_2}{\ln d_1}) = \ln d_1(1+ \frac{\ln D_2}{\ln D_1})=\ln d_1 \frac{\ln D}{\ln D_1}. 
\end{equation}
We recall that $D_1D_2=D$, the dimension of the Hilbert space of the full system. Rearranging the terms, we get:
\begin{equation}\label{ln_d1}
\ln d_1 \simeq \ln D_1 \left(\frac{\ln d_E}{\ln D}\right).
\end{equation}

This equation can be recast as:
\begin{equation}
 d_1\simeq D_1^\gamma, 
    \label{d1_eff}
\end{equation}
with $\gamma=\frac{\ln d_E}{\ln D}$ for nonintegrable (chaotic) systems.
This is the expression of $d_1$ that appears in Eq. \ref{avne1_2}. 
%For the integrable system, $d_1d_2<d_E$, as noted earlier. It is easy to check that, even for this case, one can write $d_1=D_1^\gamma$, but now $\gamma<\frac{\ln d_E}{\ln D}$. 
It is interesting to note here that, as $\gamma \to 1$, we recover the result of Page as appear in Eq. \ref{page_th}.

Since an integrable system possesses a large number of conserved quantities (represented by independent operators commuting with the Hamiltonian), the system cannot dynamically access all states in the subspace $\mathcal{H}_E$. Instead, the additional constraints restrict the dynamics to only a part or division of this subspace. Assume that there are $g$ such partitions or divisions of the subspace $\mathcal{H}_E$ due to the additional constraints. Let $d_E^{(i)}$ be the dimension of the $i$th such division. Clearly, $\sum_{i=1}^g d_E^{(i)}=d_E$. For this division, Eq. \ref{eq_de} is rewritten as $d_1^{(i)}d_2^{(i)}\simeq d_E^{(i)}$, where $d_1^{(i)}$ ($d_2^{(i)}$) is the dimension of the effective Hilbert space of the first (second) subsystem when the full system is restricted to the $i$th division of the subspace $\mathcal{H}_E$. A scaling analysis for this division yields $d_1^{(i)}\simeq D_1^{\gamma^{(i)}}$, where $\gamma^{(i)}=\frac{\ln d_E^{(i)}}{\ln D}$ (see Eq. \ref{d1_eff}).

Since $0 < d_E^{(i)} < d_E$, or equivalently $0 < \gamma^{(i)} < \tfrac{\ln d_E}{\ln D}$, the average VN entropy for a given division of the subspace $\mathcal{H}_E$, namely $\ln d_1^{(i)}$, is necessarily smaller than the maximum possible value $\ln d_1$, with $d_1$ given by Eq. \ref{d1_eff}. Consequently, because the entropy associated with each division falls below this maximum, the VN entropy $\overline{S}_1$, obtained by averaging over all divisions of $\mathcal{H}_E$ (i.e., averaged over the full subspace $\mathcal{H}_E$), is also strictly less than the maximum achievable value characteristic of a nonintegrable system. 
This result for integrable systems can be concisely expressed as follows. Corresponding to the subspace $\mathcal{H}_E$, the average VN entropy of subsystem 1 is given by $\overline{S}_1\simeq \ln{d_1}$, where $d_1=D_1^{\gamma}$ with $\gamma < \ln (d_E) / \ln (D)$. We further note that the effective value of $\gamma$ in integrable systems is not universal, since the partitioning of the subspace $\mathcal{H}_E$ is system dependent.

\subsection{Domain of validity and scope of application}
While trying to establish our main result (see Eq. \ref{d1_eff} or equivalently, Eq. \ref{s1_lndos2}), we have considered some physically motivated assumptions. It is important that we have clarity on the assumptions and the need to have them to arrive at our main result. 

In Page's case, where $\mathbb{E}(S_{1}) \simeq \ln(D_{1})$ (Eq. \ref{page_th}), all energy scales of a subsystem, and hence all of its energy eigenkets, are important and contribute equally. In contrast, when the full system is restricted to a fixed energy, the subsystem’s energy scales are not all equally relevant. Consequently, the average von Neumann entropy in our case is reduced relative to Page’s result: $\overline{S}_1 < \ln(D_{1})$. If we parameterize this entropy as $\overline{S}_1 \simeq \ln d_{1}$, it follows that \(d_{1}<D_{1}\). A natural question is then: how do we estimate the actual value of the parameter \(d_{1}\)?

To estimate $d_1$, interpreted as the dimension of the effective Hilbert space of the subsystem when the full system is energy-restricted, we proposed a scaling theory (Sec. \ref{sec2.5}). The validity of this scaling analysis can be demonstrated clearly in a particular special case, discussed below, although its actual range of applicability is expected to be broader than this illustrative scenario.

We note that for a full system constrained to have a fixed energy \(E\), if its subsystem is also settled in a fixed energy scale \(\overline{E}_1\), then \(d_{1}\) admits a clear physical interpretation. Specifically, \(d_{1}\) represents the number of subsystem eigenstates compatible with its energy \(\overline{E}_1\). In this case, \(\ln d_{1}\) is simply the Boltzmann entropy of the subsystem, which coincides with the von Neumann entropy. 
In this special scenario, the subspace dimension ($d_E$) is just the product of $d_1$ and $d_2$ (where the latter quantity is similarly defined for subsystem 2). This justifies the basis of our scaling analysis, as appears in Eq. \ref{eq_de}. 

However, the assumption that the subsystem necessarily lies at a fixed energy scale whenever the full system is energy-restricted is not universally valid. It becomes reasonable assumption under the following conditions:  (a) \(1 \ll l_{1} \ll l_{2}\), (b) the Hamiltonian is local, and  (c) \(1 \ll d_{E} \ll D\).  

Condition (c) effectively asserts that the full system is confined to a narrow energy shell. When (c) holds, conditions (a) and (b) typically ensure that the energy distribution of the first subsystem is Gaussian with a sharp peak, thereby justifying the assumption.

Another assumption we make is that the system is uniform. This uniformity allows us to assign the same value of the scaling parameter $\eta$ to both subsystems (see Sec. \ref{sec2.5}).

\section{Numerical verifications} \label{sec3}
To numerically verify the result presented in Eq. \ref{avne1_2}, with $d_1$ given in Eq. \ref{d1_eff}, we rearrange the terms and express the average VN entropy in terms of the DOS of the full system and the size of the subsystem 1.
Starting from Eq. \ref{avne1_2}, we get: 
\begin{equation}
\overline{S}_1 \simeq \ln d_1 = \left(\frac{\ln d_E}{\ln D}\right)\ln D_1.
%=\lambda\ln d_E, 
\label{s1_lndos1}
\end{equation}
Since $\frac{\ln D_1}{\ln D}=\frac{l_1}{l_1+l_2}$, we obtain the following leading term in the average entropy:
\begin{equation}
\overline{S}_1 \simeq \ln d_1 \simeq l_1\left(\frac{\ln\textrm{(DOS)}}{l_1+l_2}\right),
%=\lambda\ln d_E, 
\label{s1_lndos2}
\end{equation}
where again we used the fact that $d_E\propto \textrm{DOS}$. For an integrable system, one can again argue that the average VN entropy is less than the maximum value presented in Eq. \ref{s1_lndos2}.

\begin{figure*}
    \centering
    \includegraphics[width=12cm, height=7cm]{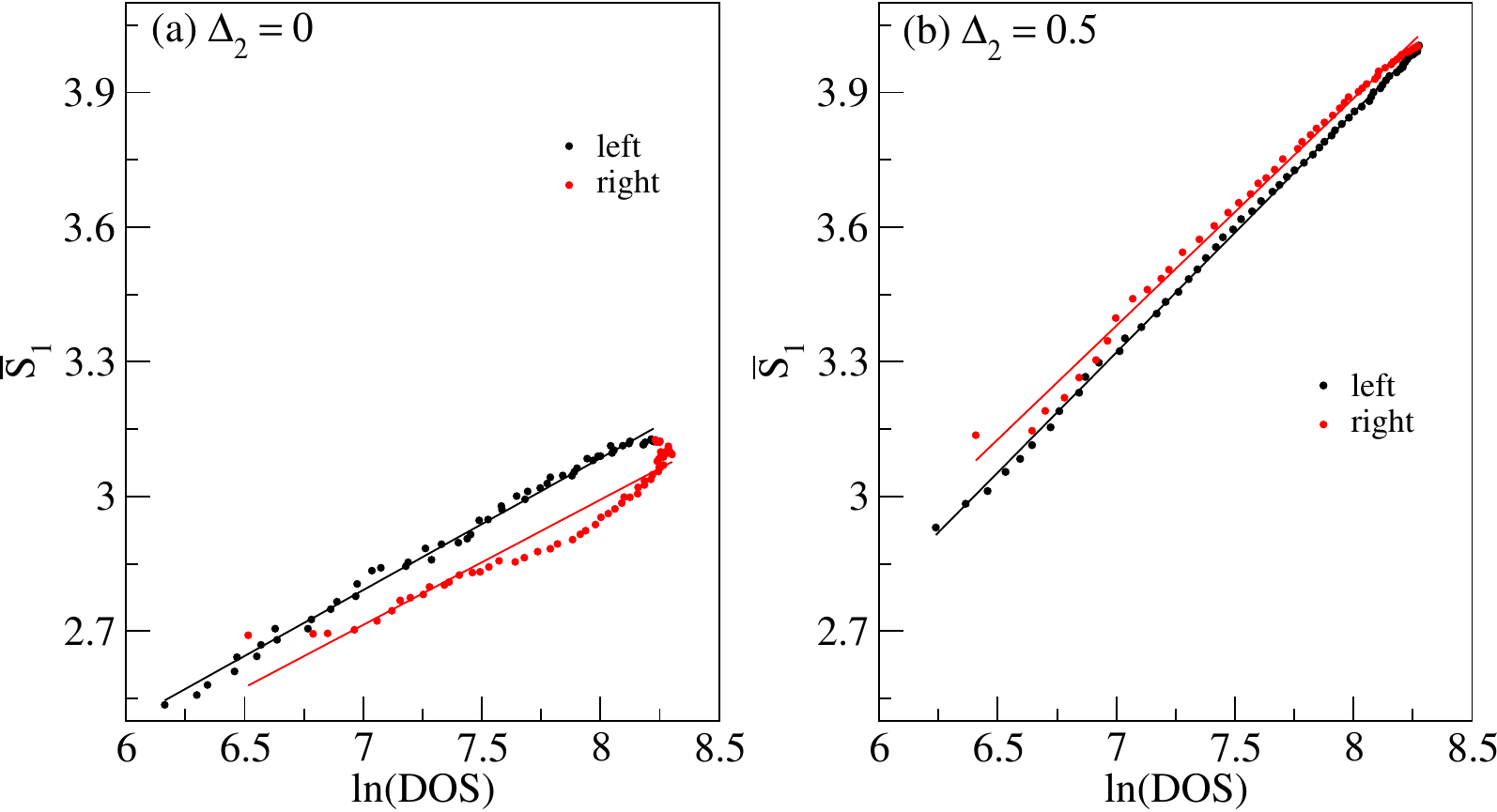}
    \caption{The average VN entropy ($\overline{S}_1$) of subsystem 1 is plotted against $\ln$(DOS) across the energy spectrum of the full system. The plots are shown separately for the left and the right halves of the spectrum. Calculations are performed with full system size $N=16$ and subsystem size $l_1=6$.}
    \label{s1_lndos_fig}
\end{figure*}

\begin{figure}
    \centering
    \includegraphics[width=7.5cm, height=6cm]{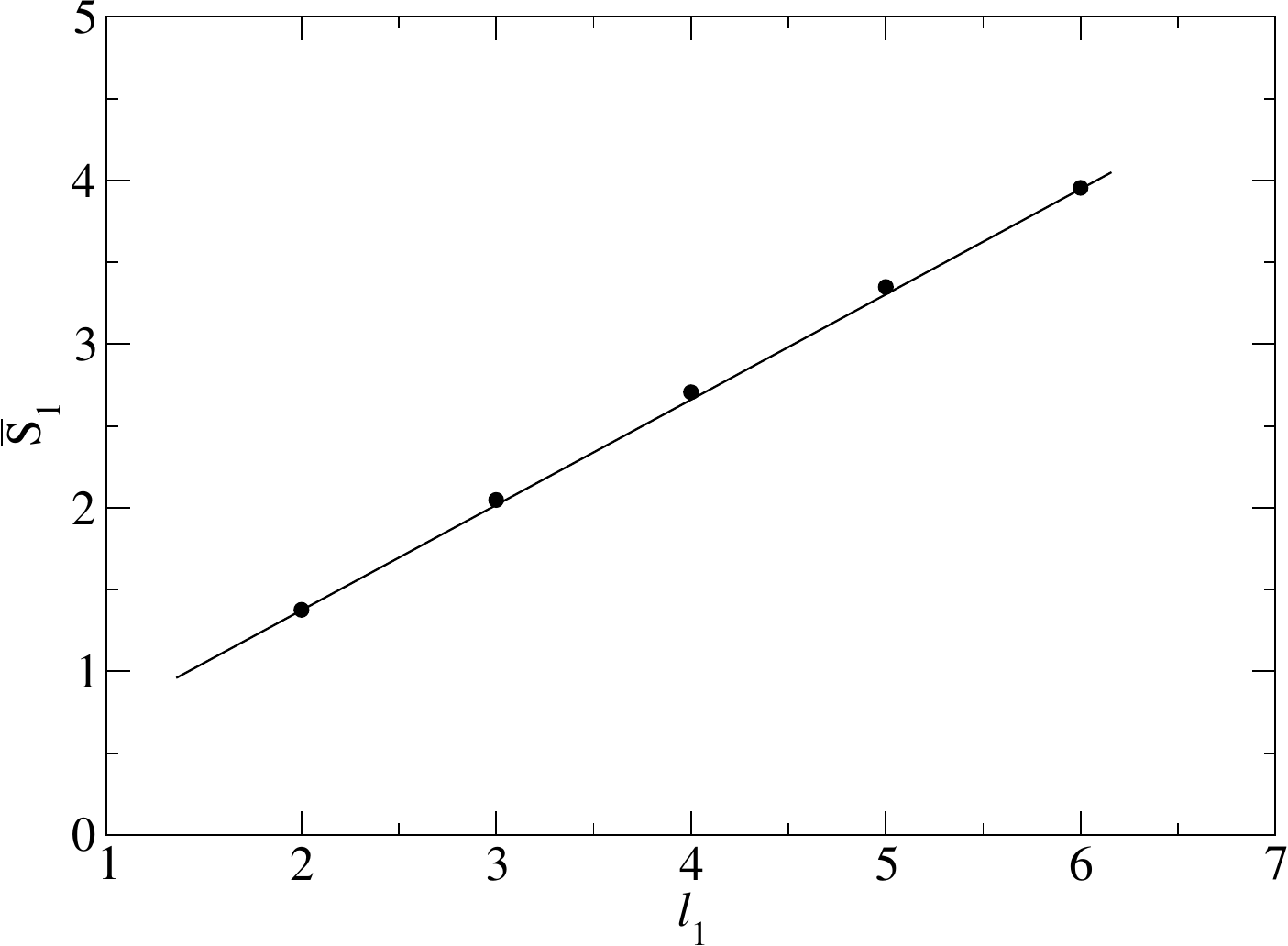}
    \caption{The subsystem average VN entropy ($\overline{S}_1$) is plotted against the subsystem size ($l_1$) for a fixed energy (corresponding to the highest DOS).}
    \label{s1_l1}
\end{figure}

We see from the above equation that, for a large system of fixed size $l_1+l_2$,  $\overline{S}_1\propto \ln\textrm{(DOS)}$ for given subsystem size $l_1$, and $\overline{S}_1\propto l_1$ for given energy (or DOS). The second relation is simply the volume-law of entropy; the coefficient is now not constant but depends on DOS (or the energy $E$). For a spin model discussed below, we numerically test our results discussed above.  

For numerical calculations, we take a spin-1/2 one-dimensional system with next-nearest neighbor interactions, as described by the following Hamiltonian:
\begin{equation}
    H=\sum_{i=1}^{N-1}\vec{S}_i\cdot \vec{S}_{i+1}+\Delta_2\sum_{i=1}^{N-2} S^z_i S^z_{i+2}.
    \label{spn_ham}
\end{equation}
This model system is integrable when $\Delta_2=0$ and is nonintegrable when $\Delta_2> 0$ \cite{Steinigeweg13}. In our calculations, we take a 16-site system as the full system (open chain) and perform all claculations in the $S_z=0$ sector ($S_z$ being the $z$ component of the total spin). 

First, we plot the VN entropy ($S_1$) of the individual eigenkets of the full system for $\Delta_2$ = 0.0, 0.2 and 0.5. We take the 6-site left-end part of the full system as subsystem 1. The result can be found in Fig. \ref{ee_lndos_eng}. In the same figure, we also show how the logarithm of DOS changes across the spectrum. Next, we evaluate $\overline{S}_1$ over a small energy interval and plot it against $\ln$(DOS) calculated at the same energy scale. The result can be found in Fig. \ref{s1_lndos_fig}, where this plot is shown separately for the left and the right halves of the spectrum. The linearity of these plots verifies that $\overline{S}_1\propto\ln{\textrm{(DOS)}}$ for given subsystem size $l_1$ (see Eq. \ref{s1_lndos2}). We also note from the figure that the slope is largest for the nonintegrable case ($\Delta_2=0.5$) but is smaller than that for the integrable case ($\Delta_2=0.0$), as expected from our earlier discussion. It may be noted that, for the integrable case, the linear fit is not always as good as the fit for the nonintegrable case. This shows that the proportionality constant in $\overline{S}_1\propto\ln{\textrm{(DOS)}}$ can depend on the energy scale for integrable systems. 

Last, we verify the volume law for this spin model. We plot $\overline{S}_1$ against the subsystem size $l_1$ (for the fixed energy where the DOS is maximum). As predicted by Eq. \ref{s1_lndos2}, we see a linear plot in Fig. \ref{s1_l1} - implying that $\overline{S}_1 \propto l_1$. For our calculations, we choose $\Delta_2=0.5$ (where the system is nonintegrable). 

\section{Conclusion}\label{sec4}
Although the average von Neumann (VN) entropy for Haar-random pure states is well known (Page's result), it is important to know the average for physically relevant subspaces. In this Letter, we analyze the average von Neumann (VN) entropy corresponding to an energy scale (for generic quantum local Hamiltonians). We find that the average entropy of a part of a system (here, subsystem 1) is $\overline{S}_1 \simeq \ln{d_1}$, where $d_1=D_1^\gamma$ ($D_1$ is the Hilbert space dimension of the subsystem 1). We argue that $\gamma=\frac{\ln{d_E}}{\ln{D}}$ for nonintegrable (chaotic) systems, and $\gamma<\frac{\ln{d_E}}{\ln{D}}$ for integrable systems, where $D$ is the Hilbert space dimension of the full system and $d_E$ is the subspace dimension corresponding to the energy scale. This result can be reinterpreted as the volume-law of entropy where the coefficient of volume is dependent on the density of states (DOS) for nonintegrable system and the coefficient is less than the maximum possible value for integrable system. We substantiate our results by numerical analysis of a one-dimensional spin chain. Our results provide a complementary understanding of quantum thermalization which is traditionally explained in the frameworks such as canonical typicality and the Eigenstate Thermalization Hypothesis (ETH).

\section{Acknowledgement}
SS thanks S. Aravinda for useful discussions.

%\bibliographystyle{elsarticle-harv}
%\bibliography{manuscript}

\bibliographystyle{plain}        % Include this if you use bibtex 
\bibliography{manuscript_ref}

\end{document}